\documentstyle[twoside,fleqn,espcrc2,epsf]{article}

\newcommand{\AmS}{{\protect\the\textfont2
  A\kern-.1667em\lower.5ex\hbox{M}\kern-.125emS}}

\title{Effects of Nonperturbative Improvement in Quenched Hadron 
       Spectroscopy\thanks{Talk presented by T. Mendes}}

\author{A. Cucchieri, T. Mendes and R. Petronzio\address{Gruppo APE,
        Dipartimento di Fisica,
        Universit\`a di Roma ``Tor Vergata'' and INFN -- Sezione di Roma 2,
        Via della Ricerca Scientifica 1,
        00133 Roma, ITALY}}
\begin{document}

\begin{abstract}
We discuss a comparative analysis of unimproved and
nonperturbatively improved quenched hadron spectroscopy, on a set
of 104 gauge configurations, at beta=6.2. We also present here our 
results for meson decay constants, including the constants $f_D$ 
and $f_{D_s}$ in the charm-quark region.
\end{abstract}

\maketitle

We report on our simulations of hadron spectroscopy
using the improved action
\begin{eqnarray}
S &=& S_G \,+\, S_W \nonumber \\
             & & \quad \,+\, c_{SW}\,a^5\,\frac{i}{4}
\sum_x {\overline \Psi}(x)\,\sigma_{\mu\nu}{\hat F}_{\mu\nu}\Psi(x)
\;\mbox{,}
\nonumber
\end{eqnarray}
with $c_{SW}$ determined nonperturbatively by the Alpha collaboration.
(For details see \cite{previous1,previous2} and references thereof.)

Our runs consist of 104 quenched configurations, at $\beta=6.2$
and volume $24^3\times48$. We consider 7 values of the hopping 
parameter $\kappa$, and all their nondegenerate flavor combinations. 
We run at the improved value $c_{SW}= 1.61375065$ and also, for comparison,
at $c_{SW}=0$. We have considered point-like sources for the inversion, 
and our analysis is done using single-elimination jack-knife.

We are particularly interested in determining the quark-mass dependence 
of our observables. To this end, the usual bare quark mass
$\,m_q\,=\, 1/2\,(1/\kappa\,-\,1/\kappa_c)\,$ is considered in the 
{\bf unimproved} case, whereas for the {\bf improved} case we use the 
improved mass, given by
$$
{\widetilde m_q}\;\equiv\;\frac{m^R}{Z_m}\;=\;
m_q\,(1\,+\,b_m\,m_q)
$$
with the nonperturbative value $b_m\,=\,-0.62(3)$.
We generally take functions of the symmetric mass average
(i.e.\ a single variable) for our plots and chiral fits.
In the case of mesons, for example, we take
$$
{\widetilde m_q}(\kappa_1,\kappa_2)\;\equiv\;
\frac{{\widetilde m_q}(\kappa_1)\,+\,{\widetilde m_q}(\kappa_2)}{2}\;,
$$
where $\kappa_1$ and $\kappa_2$ correspond to two quark flavors.
We find that for some observables this may not be a good approximation 
in the region of heavier quark masses. We mention two such cases:
1) octet (spin-$1/2$) baryon masses, which are not expected to be functions
of a flavor-symmetric combination of the bare quark masses (nevertheless,
we have shown in \cite{previous1} that this effect is considerably attenuated 
when one employs the improved bare quark mass ${\widetilde m_q}$); and 2)
pseudoscalar meson decay constants, for which we find considerable
deviations from the ``single-variable'' curve in the charm-quark region. 
In this paper we discuss the latter case in some detail.

In order to determine $\kappa_c$ we use
the unrenormalized current quark mass, defined as
$$ \hskip -3mm m_{WI} \; \equiv \;
\frac{ \langle \partial_{\mu} \{ {A_{\mu}^{(bare)}} \,+\,
c_A \, a \, \partial_{\mu}{P^{(bare)}}\}{\cal O} \rangle}
{2 \langle{P^{(bare)}}{\cal O} \rangle}\;.$$
We use {$M_{PS}^2/M_V^2$ to get the strange-quark mass
in lattice units $m_s$, and we then determine the inverse lattice spacing
$a^{-1}$ by imposing $a^{-1}\,M_V(m_s/2)\,=\,M_{K^{*}}$. 
We obtain the following results:
\begin{center}
\addtolength{\tabcolsep}{-0.5mm}
\begin{tabular}{ccc}
\hline
   & unimproved & improved \\
\hline
{$\kappa_c$} & 0.153230(15) & 0.135828(8) \\
{$a^{-1}$} & 2.94(10) GeV & 2.56(10) GeV \\
\hline
\end{tabular}
\end{center}

\vskip 3mm
From our data for light-meson observables (see \cite{previous1,previous2}) 
we do not see a large effect of improvement. In particular, known problems 
of the quenched approximation are observed also in the improved
results: we see discrepancy in the value of the $J$ variable when 
compared with the experimental value, we get inconsistent
estimates for the inverse lattice spacing depending on the quantity used
as experimental input, and the hyperfine splitting between 
vector and pseudoscalar mesons, which is experimentally constant,
drops monotonically as we move to heavier quark masses (see Fig.\ \ref{hfs}). 
\begin{figure}
\begin{center}
\protect\vspace{-0.7cm}
\epsfxsize = 0.50\textwidth
\protect\hspace*{-0.35cm}
\leavevmode\epsffile{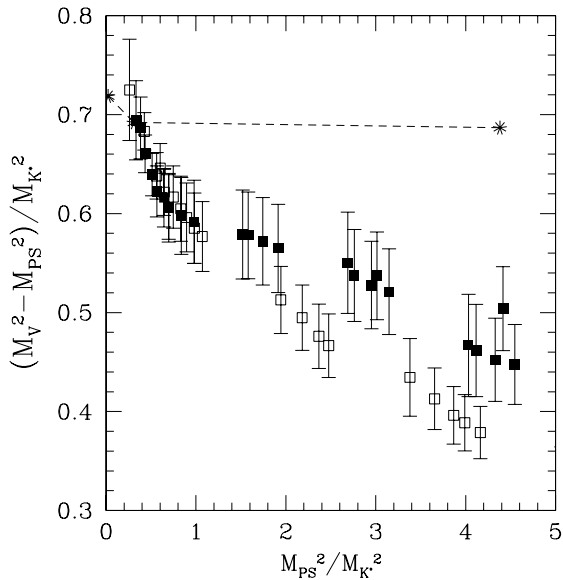}
\protect\vspace{-0.9cm}
\caption{Hyperfine splitting in units of the $K^{*}$ meson mass.
         Experimental points (in the light-, strange- and charm-quark
         regions) are connected by a dashed line,
         squares and filled squares represent the unimproved and 
         the improved case, respectively.}
\label{hfs}
\protect\vspace{-1.2cm}
\end{center}
\end{figure}
We point out, in any case, that a slight improvement is observed
in the spread of lattice spacings coming from different physical inputs
(see \cite{previous2}), and that, as seen in Fig.\ \ref{hfs}, although 
the improved and unimproved values for the hyperfine splitting
are identical within error bars at lighter quark masses,
the deviation from the experimental value is noticeably smaller 
in the improved case for heavier quark masses. 

In Table \ref{light_decay} we present our data for the light-meson decay 
constants.  The difference ``unimproved $\,-\,$ improved'' is computed
from each jack-knife cluster and subsequently averaged.
We see a slight improvement in the case of $f_{\pi}$ and no improvement
within error bars for the other cases.
(In our evaluations for decay constants we use nonperturbative 
results for all the improvement coefficients and renormalization
constants, except for the unimproved renormalization constant $Z_A$
and the improvement coefficient $b_A$, for which we used tadpole-improved
results.)
\begin{table}
\protect\vspace{-0.3cm}
\addtolength{\tabcolsep}{-1mm}
\begin{center}
\caption{Results for light-meson decay constants. The
         numbers in the first two rows of data are given in MeV;
         their errors do not include the 4\% uncertainty 
         coming from $a^{-1}$.}
\begin{tabular}{ccccc}
\hline
 & EXP & UNIMP & IMP & DIFF \\
\hline
$f_{\pi}$ & 131 & 137(10) & 132(8) & 5(5)\\
$f_K$ & 160 & 154(7) & 152(6) & 2(3) \\
$f_{\pi}/M_{\rho}$ &  0.17 & 0.172(15) & 0.167(15) & 0.005(11) \\
$f_K/f_{\pi}$ & 1.22 & 1.122(36) & 1.149(29) & -0.027(26) \\
${f_{\rho}}^{-1}$ & 0.28 & 0.302(13) & 0.255(18) & 0.047(15) \\
\hline
\label{light_decay}
\end{tabular}
\protect\vspace{-1.2cm}
\end{center}
\end{table}
\begin{figure}
\begin{center}
\protect\vspace{-0.7cm}
\epsfxsize = 0.48\textwidth
\protect\hspace*{-0.1cm}
\leavevmode\epsffile{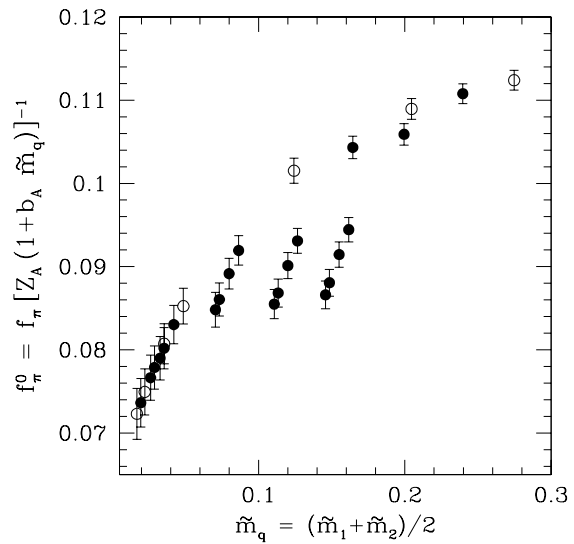}
\protect\vspace{-1.2cm}
\caption{Pseudoscalar-meson decay constant for the improved case.
         Open circles correspond to diagonal flavor combinations
         (${\widetilde m_1}={\widetilde m_2}$).}
\protect\vspace{-1.2cm}
\end{center}
\end{figure}

For the case of the pseudoscalar decay constant, as seen in Fig.\ 2
for the improved case, the data points corresponding to nondegenerate
flavor combinations (filled circles) deviate considerably
from the diagonal-flavor curve, and a fit using
a function of the average quark mass 
$\,({\widetilde m_1}+{\widetilde m_2})/2\,$ is very poor.
This effect is not
seen in the light- and strange-quark regions, but plays a crucial role
in the determination of decay constants in the charm-quark region.
We have considered in this mass region a more general fit, 
allowing for a term of the form 
$\,({\widetilde m_1}^2+{\widetilde m_2}^2)/2\,$ in addition to the
linear and quadratic terms in the average mass. This
is still a symmetric function of the two quark masses ${\widetilde m_1}$
and ${\widetilde m_2}$.
This fit is able to describe the numerical data with very good accuracy,
and, in the light-quark region, gives results consistent with the ones 
of the single-variable fit.
We give our results for the constants $f_D$ and $f_{D_s}$
in Table \ref{fD}. The improved data show very good agreement with
the average of world lattice data given in Reference \cite{fD_ref}.
\begin{table}
\begin{center}
\protect\vspace{-0.7cm}
\caption{Pseudoscalar-meson decay constants in the charm-quark region.
         The values in the first two rows of data are given in MeV;
         their errors do not include the 4\% uncertainty 
         coming from $a^{-1}$.}
\addtolength{\tabcolsep}{-1mm}
\begin{tabular*}{6.5cm}{cccc}
\hline
 & UNIMP & IMP & DIFF \\
\hline
$f_D$ & 208(3) & 201(4) & 7(2) \\
$f_{D_s}$ & 220(3) & 222(4) & -2(2) \\
$f_{D_s}/f_D$ & 1.057(3) & 1.084(5) & -0.027(3) \\
\hline
\end{tabular*}
\label{fD}
\end{center}
\protect\vspace{-1cm}
\end{table}
We note that a fit done using only diagonal points 
(${\widetilde m_1}={\widetilde m_2}$) gives fit coefficients
that are consistent with our more general fit. However,
using the ``diagonal'' fit and the average-mass approximation
to compute $f_D$ and $f_{D_s}$ leads to highly overestimated
values: we get, in the improved case $f_D=265$ MeV and
$f_{D_s}=275$ MeV.

\vskip 3mm
Finally, we have found in \cite{previous1,previous2} that two 
different determinations of the strange-quark mass are in much 
better agreement in the improved than in the unimproved case, 
and that it is very important that improved quark masses 
${\widetilde m_q}$ be used for the determination based on the 
lattice ratio of charm-quark over strange-quark masses. 
Another sensible effect of improvement is seen in the baryon mass 
values, given in Table \ref{bar_masses} in MeV, together with their 
jack-knife differences. 
We also include here an APE plot for the decuplet baryons (Fig.\ 
\ref{APE}), from 
which we see very clear agreement in the improved case between 
experimental and numerical points, while the unimproved points show 
the wrong slope.
Within our error bars, we see no effect of quenching for these masses.
\begin{table}
\begin{center}
\protect\vspace{-0.7cm}
\caption{Baryon mass values in MeV and comparison with experiment.}
\addtolength{\tabcolsep}{-1mm}
\begin{tabular*}{7.5cm}{ccccc}
\hline
 & EXP & UNIMP & IMP & DIFF \\
\hline
$M_N$ & 939 &
1055(97) & 953(117) & 102(117) \\
$M_{\Lambda}$ & 1115.7 &
1205(80) & 1148(76) & 57(74) \\
$M_{\Delta}$ & 1232 &
1500(150) & 1265(113) & 235(109) \\
$M_{\Sigma-\Lambda}$ & 73.7 &
49(23) & 70(29) & -22(29) \\
$M_{\Delta-N}$ & 293 &
336(59) & 297(80) & 39(62) \\
\hline
\end{tabular*}
\label{bar_masses}
\end{center}
\protect\vspace{-0.8cm}
\end{table}

\begin{figure}
\begin{center}
\protect\vspace{-0.7cm}
\epsfxsize = 0.51\textwidth
\protect\hspace*{-0.5cm}
\epsffile{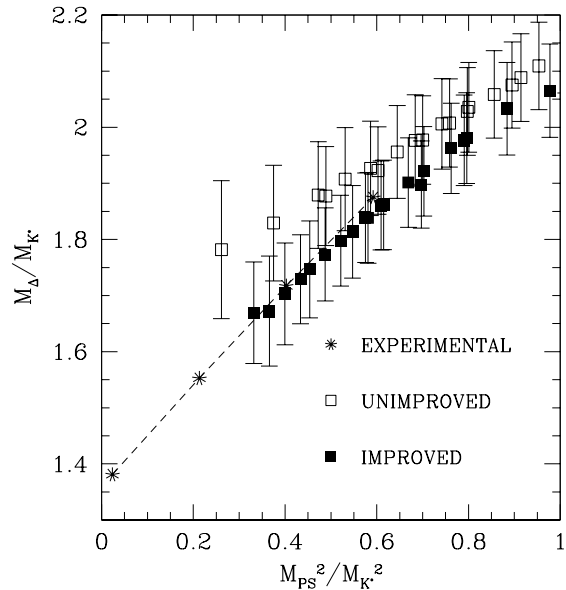}
\protect\vspace{-0.8cm}
\caption{APE plot for decuplet baryons.}
\label{APE}
\end{center}
\protect\vspace{-0.7cm}
\end{figure}


\begin{thebibliography}{9}
\bibitem{previous1} A. Cucchieri, M. Masetti, T. Mendes and R. Petronzio,
                    Phys.\ Lett.\  B422, 212 (1998);
\bibitem{previous2} A. Cucchieri, T. Mendes and R. Petronzio,
                    J.\ High Energy Phys.\ 05, 006 (1998).
\bibitem{fD_ref}    J.M. Flynn and C.T. Sachrajda, {\em Heavy Quark
                    Physics from Lattice QCD}, {\tt hep-lat}/9710057.
\end{thebibliography}
\end{document}